\documentclass{eas}
\usepackage{graphicx}
\usepackage{amsmath}
\usepackage{amsfonts}
\usepackage{amssymb}

\def\corot{\small CoRoT}
\def\helas{\small HELAS}
\def\esta{{\small ESTA}}

\def\ASTEC{{\small\bf ASTEC}}
\def\CESAM{{\small\bf CESAM}}
\def\CLES{{\small\bf CL\'ES}}
\def\FRANEC{{\small\bf FRANEC}}
\def\TGEC{{\small\bf TGEC}}
\def\GARSTEC{{\small\bf GARSTEC}}

\def\astec{{\small ASTEC}}
\def\cesam{{\small CESAM}}
\def\cles{{\small CL\'ES}}
\def\franec{{\small FRANEC}}
\def\tgec{{\small TGEC}}
\def\garstec{{\small GARSTEC}}

\def\etal{{\em et al.}}

\def\EAS{these proceedings, EAS Publication Series}
\def\workshop{Joint \helas/\corot\ Workshop, Porto, Portugal, at {\tt\small http://www.astro.up.pt/investigacao/conferencias/hce2006/}}

\newcommand{\msol}{\mbox{${\mathrm M}_{\odot}$}}

\begin{document}

%%-----------------------------
%%      the top matter
%%-----------------------------
\title{Microscopic Diffusion in Stellar Evolution Codes: First Comparisons results of ESTA-Task~3} 
\runningtitle{Diffusion in Stellar Evolution Codes: First Comparisons of ESTA-Task~3}
\author{Y.~Lebreton}\address{Observatoire de Paris, GEPI, CNRS UMR 8111, Meudon, France}
\author{J.~Montalb\`an}\address{Institut d'Astrophysique et de G\'eophysique, Universit\'e de Li\`ege, Belgique}
\author{J.~Christensen-Dalsgaard}\address{Institut for Fysik og Astronomi, Aarhus Universitet, Denmark}
\author{S.~Th\'eado}\sameaddress{2}
\author{A.~Hui-Bon-Hoa}\address{LATT, Observatoire de Midi-Pyr\'en\'ees, CNRS UMR 5572, Toulouse, France}
\author{M.~J.~P.~F.~G.~Monteiro}\address{Centro de Astrof\'{\i}sica da Universidade do Porto, and Departamento de Matem\'atica Aplicada, Faculdade de Ci\^encias da Universidade do Porto, Portugal}
\author{S.~Degl'Innocenti}\address{Dipartimento di Fisica, Universit\`a di Pisa, Italy}
\author{M.~Marconi}\address{INAF-Osservatorio Astronomico di Capodimonte, Napoli, Italy}
\author{P.~Morel}\address{Observatoire de la C\^ote d'Azur, Cassiop\'ee, CNRS UMR 6202, Nice, France}
\author{P.G.~Prada Moroni}\sameaddress{6}
\author{A.~Weiss}\address{Max-Planck-Institut f\"ur Astrophysik, Garching, Germany}

\begin{abstract}
We present recent work undertaken by the \textit{Evolution and Seismic Tools Activity} (\esta) team of the \corot\ \textit{Seismology Working Group}. The new \esta-Task~3 aims at testing, comparing and optimising stellar evolution codes which include microscopic diffusion of the chemical elements resulting from pressure, temperature and concentration gradients. The results already obtained are globally satisfactory, but some differences between the different numerical tools appear that require further investigations. 
\end{abstract}
\maketitle
%%-----------------------------
%%      your text
%%-----------------------------
\section{Introduction}

In previous papers we have presented the work undertaken by the 
\textit{Evolution and Seismic Tools Activity} (\esta) team of the \corot\ \textit{Seismology Working Group} (see  Monteiro \etal~\cite{monteiro06} and references therein).
In this activity, our main goal is to test, compare and optimise numerical tools which will be used to model the internal structure and evolution of the \corot~target stars and to calculate their oscillation properties.

\esta-Task~1 is now finished, it has concentrated on the comparison of standard stellar models coming from seven stellar evolution codes in the range of mass (from $0.9$ to $5.0$\ \msol) and evolution stages (from the pre-main sequence to the subgiant branch) to be covered by the \corot~main targets. Task~2, still underway, has concentrated on seismic codes (see Moya \cite{moya07}). The results so far obtained in Tasks~1 and 2 are quite satisfactory, showing minor differences between the different numerical tools provided the same assumptions on the physical parameters are made. These first comparison steps have given us confidence on the numerical tools that will be available to interpret the future \corot\ seismic data.

We present here the new \esta-Task~3 devoted to the comparison of stellar models taking into account microscopic diffusion of chemical elements resulting from pressure, temperature and concentration gradients (see Thoul \& Montalb\`an \cite{tm07}). In this step, we do not take into account diffusion due to the radiative forces, nor the extra-mixing of chemical elements due to differential rotation or internal gravity waves (see Alecian \cite{alecian07}; Mathis \cite{mathis07}; Zahn \cite{zahn07}).

Evolution models of 1.0, 1.2 and 1.3 \msol~have been calculated by the \esta\ group with different stellar evolution codes up to the subgiant branch. We present the first comparisons of those stellar models at particular stages of the evolution. Detailed comparisons of some of the codes and discussions of models are presented by Mont\`alban \etal\ (\cite{mtl07}), Christensen-Dalsgaard(\cite{jcd07}), Marconi (\cite{marconi07}) and Christensen-Dalsgaard \& Di Mauro (\cite{jcddm07}). 

\section{Specifications of \esta-Task~3}

\subsection{Input physics}

The physical assumptions proposed as the reference for the comparisons are the same as used for Task~1 and no overshooting and are described in Monteiro \etal~(\cite{monteiro06}). Regarding diffusion, we focus on helium and heavy element diffusion due to pressure, temperature and concentration gradients.

As reviewed by Thoul \& Montalb\`an (\cite{tm07}), two approaches to obtain the diffusion equation from the Boltzmann equation for binary or multiple gas mixtures can be followed: one is based on the Chapman-Enskog theory (Chapman \& Cowling \cite{chapman70}, hereafter CC70) and the other on the resolution of the Burgers equations (Burgers \cite{burgers69}, hereafter B69). In both methods, approximations have to be made to derive the various coefficients entering the diffusion equations, in particular the diffusion velocities which are written as a function of the collision integrals. In the stellar evolution codes which have participated to Task~3, either the CC70 or the B69 approach has been used (see Section~\ref{subsec:models} below).

%=====================
\begin{table}[htb]
\vskip -5pt
\caption{Specification of the models. Left: The three cases with corresponding masses and initial chemical composition. Right: The three evolutionary stages examined for each case. Phases A and B are respectively in the middle and end of the M--S stage. Phase C is on the subgiant branch. $X_c$ denotes the central H abundance in mass fraction and the He core is defined as the region of the star where the H abundance $X$ is lower than 0.01.}%\vspace{1em}  
\begin{center}
\begin{tabular}{llll}
\hline\\[-10pt]
\begin{small}case\end{small} & $\frac{M}{M_\odot}$ & $Y_0$ & $Z_0$ \\[4pt]\hline
3.1& 1.0 & 0.27 & 0.017 \\
3.2& 1.2 & 0.27 & 0.017 \\
3.3& 1.3 & 0.27 & 0.017 \\
\hline
\end{tabular}
\hspace*{2cm}
\begin{tabular}{lll}
\hline
\begin{small}phase\end{small} & $X_c$ & $M_{\mathrm{He\ core}}$ \\
\hline
A& 0.35 & - \\
B& 0.01 & - \\
C& 0.00 & 0.05$M_{\mathrm{star}}$ \\
\hline
\end{tabular} 
\renewcommand{\arraystretch}{1.2}
\label{table:specif}
\end{center}
\vskip -5pt
\end{table}
%===============

\subsection{Cases for model comparison}

During the $10^{\mathrm{th}}$ \corot\ Week in Nice we defined 3 cases (i.e. three values of the stellar mass) for the models to be compared under Task~3 (Lebreton \cite{lebreton06}). These cases are presented in Table~\ref{table:specif}. We chose rather low values of the masses (i.e. $M<1.4M_\odot$) for which diffusion resulting from radiative forces can be neglected. Furthermore, this avoids the problems occurring at higher masses where the use of microscopic diffusion alone produces a very important depletion of helium and heavy elements at the surface (and a concomitant increase of the hydrogen content) and in turn requires to invoke other mixing processes to control the gravitational settling (see for instance Turcotte \etal\ \cite{turcotte98}). For each case, models at different evolutionary stages have been considered. We focused on three particular evolution stages : middle of the main-sequence (M--S), end of the M--S, subgiant branch.

\subsection{Participating stellar internal structure and evolution codes}
\label{subsec:models}

Up to now six stellar evolution codes have been involved in Task~3. We give below brief information on the way diffusion has been implemented in each of them. More details can be found in Monteiro \etal~(\cite{monteiro06}) and references therein as well as in the presentations made during the Joint \helas-\corot\ Workshop which can be downloaded at the Web site {\tt\small http://www.astro.up.pt/investigacao/conferencias/hce2006/}.

\begin{itemize}

\item \ASTEC\ -- {\em Aarhus Stellar Evolution Code}:
In \astec\ diffusion is treated according to the simplified Michaud \& Proffitt formalism (\cite{mp93}, hereafter MP93) based on the B69 approach in which the heavy elements are treated as trace elements (see Christensen-Dalsgaard \cite{jcd07}). Models with either pure He diffusion or He-Z diffusion have been calculated (hereafter \astec-He and \astec-He-Z models). In the latter case all heavy elements are represented by ${\mathrm {^{16}O}}$. Models have 1242 mesh points and the number of time steps to reach phase C is in the range 200-2000 depending on the model.
  
\item \CESAM\ -- {\em Code d'\'Evolution Stellaire Adaptatif et Modulaire}:  
Two formalisms for diffusion have been implemented in the \cesam2k code: the simplified MP93 formalism and a general formalism based on the resolution of Burger's equations and the Paquette \etal~(\cite{paquette86}) collision integrals (Morel \cite{morel97}; Lebreton \cite{lebreton07}). In the present models He and seven heavy elements have been followed explicitly and the ionisation degree of each species has been  calculated. We consider three series of models: \cesam-V1-MP models where the MP93 formalism has been used, \cesam-V1-B69 models where the Burgers equations are solved and models calculated with \cesam-V2, the last version of \cesam2k which is still under development. \cesam-V1 models have from 2700 to 3000 mesh points and the number of time steps to reach phase C is in the range 1000-2000 depending on the model while \cesam-V2 models have from 800 to 1000 mesh points and take 100-150 time steps to reach phase C.
  
\item \CLES\ -- {\em Code Li\'egeois d'\'Evolution Stellaire}:
The most advanced version of the \cles\ code can compute the abundance variations due to microscopic diffusion for a dozen species. In the present models electrons and three species (H, He and a mean Z, all assumed to be fully ionised) have been followed. The diffusion velocities are computed following the theory developed by Thoul \etal\ (\cite{tbl94}, hereafter TBL94) which is based on the B69 approach (see Th\'eado \cite{theado07}). Models with either pure He diffusion or He-Z diffusion have been calculated (hereafter \cles-He and \cles-He-Z models). Models have about 2400 mesh points and the number of time steps to reach phase C is between 1000 and 1500.
  
\item \FRANEC\ -- {\em Pisa Evolution Code}:
Diffusion is implemented following the TBL94 theory (B69's approach). The diffusion of He and of eight heavy elements is explicitly treated (see Marconi \cite{marconi07}). Models have between 400 and 2000 mesh points and the number of time steps to reach phase C is around 1400.

\item \GARSTEC\ -- {\em Garching Evolution Code}:
In present models, diffusion is calculated following the TBL94 theory (B69's approach). Either the diffusion of He or the diffusion of all elements with diffusive speed taken that of ${\mathrm {^{56}Fe}}$ are taken into account (Weiss \cite{weiss05}). Models have between 1200 and 2000 mesh points and the number of time steps to reach phase C is around 200. In \garstec\ there is also the option (not used here) to derive diffusion constants from Paquette \etal's (\cite{paquette86}) collision integrals with quantum corrections from Schlattl \& Salaris (\cite{schlattl03}).

\item \TGEC\ -- {\em Toulouse-Geneva Evolution Code}:
Diffusion is treated following the CC70 approach with collision integrals derived from Paquette \etal\ (\cite{paquette86}). The diffusion of He and eight heavy elements, assumed to be fully ionised, is explicitly considered (see Hui-Bon Hoa \cite{hui07}). Models have between 900 and 1000 mesh points.
\end{itemize}

\begin{figure}[htb]
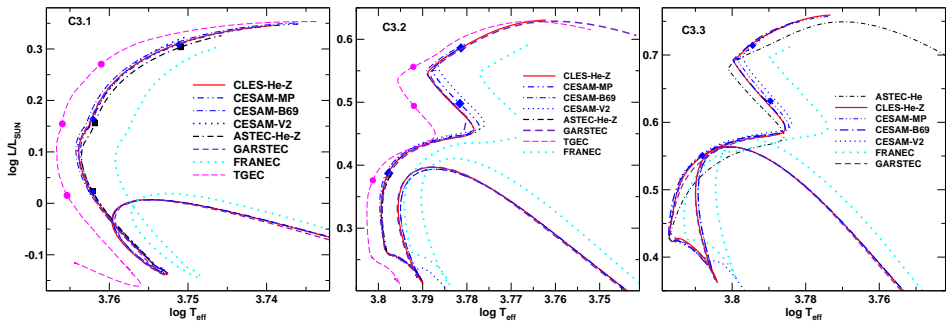

\begin{center}
\resizebox*{\hsize}{!}{\includegraphics*{./figs/HR_C31}\hspace*{0.2cm}\includegraphics*{./figs/HR_C32}\hspace*{0.cm}\includegraphics*{./figs/HR_C33}}
\caption{Evolutionary tracks and/or location of the target models A, B, C in the H--R diagram obtained for the 3 cases with the codes \astec-He (black, small dot-dash), \astec-He-Z (black, dash-dash-dot), \cesam-V1-MP (blue, dash-dot-dot), \cesam-V1-B69 (blue, large dot-dash), \cesam-V2 (blue, dots), \cles-He-Z (red, solid), \franec\ (cyan, dot), \garstec\ (indigo, small dash) and \tgec\ (magenta, large dash).}
%\vspace*{1cm}
\label{fig:hr}
\end{center}
\end{figure}

\section{Models comparison}
\subsection{H--R diagram}

Fig.~\ref{fig:hr} shows, for the different codes, the evolutionary tracks in the H--R diagram for the three cases considered and/or the location of the target models A, B, C. Note that all tracks have not been provided for each case and that in the case of \astec\ we chose to plot models including diffusion of helium and metals when available, rather than models including He diffusion only. Tracks obtained with the \astec-He-Z, \cesam-V1 (MP or B69), \cles\ and \garstec\ codes are very close. For these codes the differences in luminosity, radius and effective temperature generally remain well below 1 per cent except for Case~3.1B where the maximum difference in luminosity is close to 2 per cent. For the \franec\ and \tgec\ models and to a lesser extent \cesam-V2 models, differences with the other models amount to several per cents (with a maximum of about 25 per cents for Case~3.1C), that is at the same level or even larger than what we had found in Task~1 where the comparisons had covered a larger range of masses, chemical compositions and evolutionary stages but without diffusion. This could indicate that there remain small differences in the input physics of the \cesam-V2, \tgec\ and \franec\ models (opacities, equation of state, nuclear reaction rates, etc.) with respect to those specified for the comparisons. Although many efforts have been made in Task1 (see for instance Degl'Innocenti \& Marconi \cite{scilla05}), we have not yet fully traced their cause.

%This could indicate that some of the input physics of the \cesam-V2, \tgec\ and \franec\ models (for instance opacities, equation of state or nuclear reaction rates) still do not follow exactly the specifications proposed for the comparisons.

\begin{figure}[htb]
\begin{center}
\resizebox*{\hsize}{!}{\hspace*{-0.cm}\includegraphics*{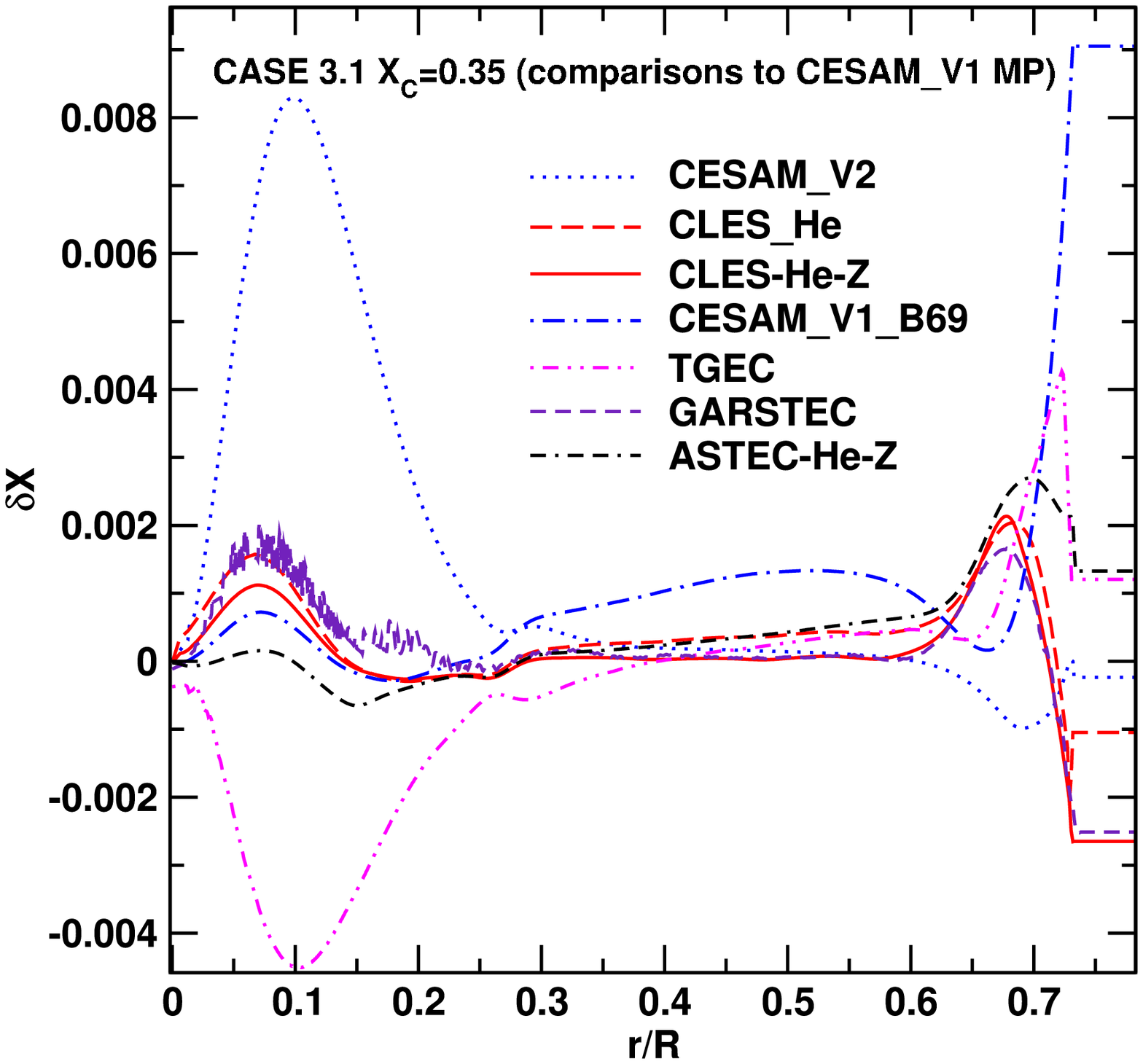}\hspace*{0.4cm}\includegraphics*{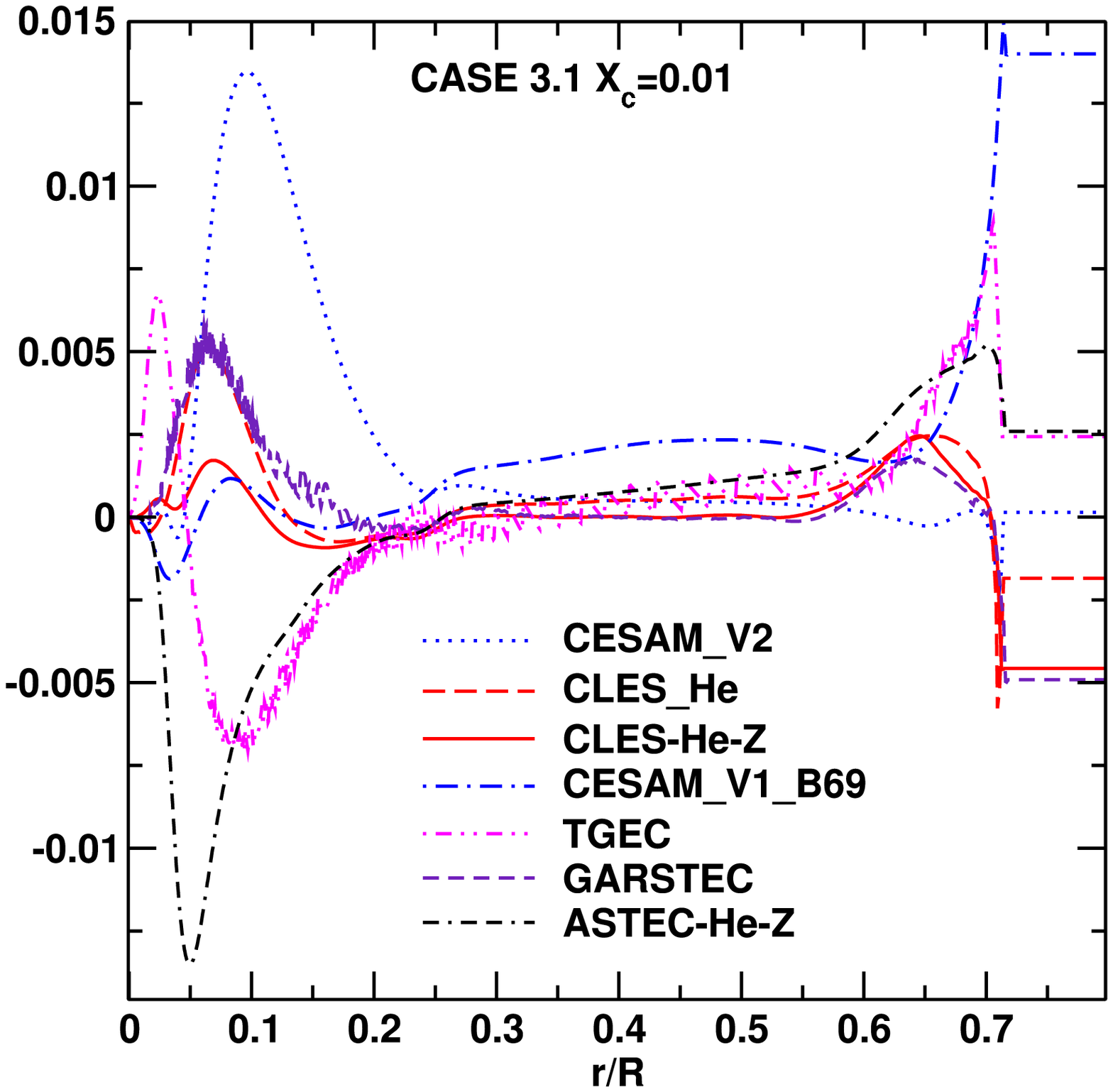}\hspace*{0.7cm}\includegraphics*{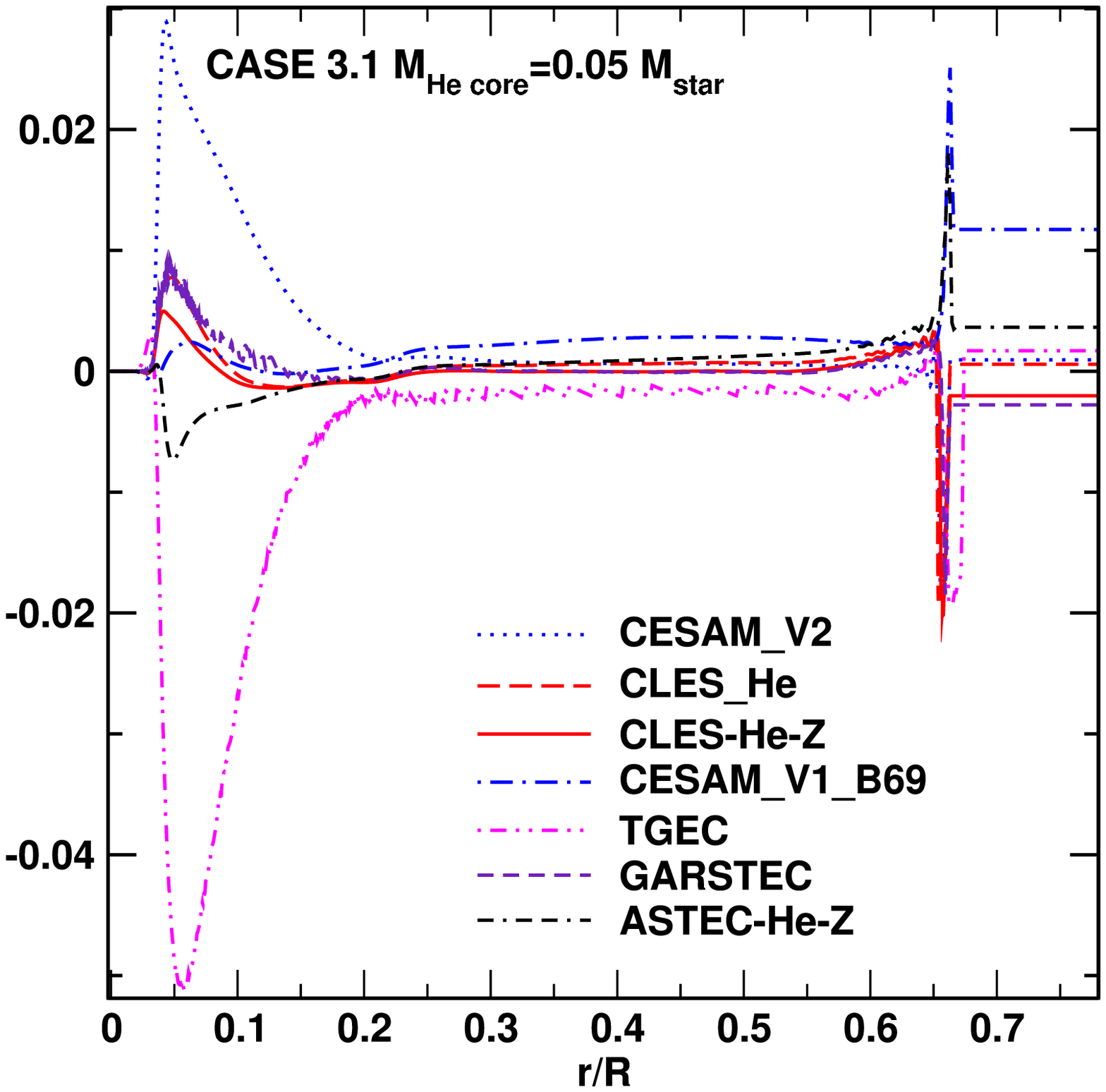}}
\resizebox*{\hsize}{!}{\hspace*{-0.cm}\includegraphics*{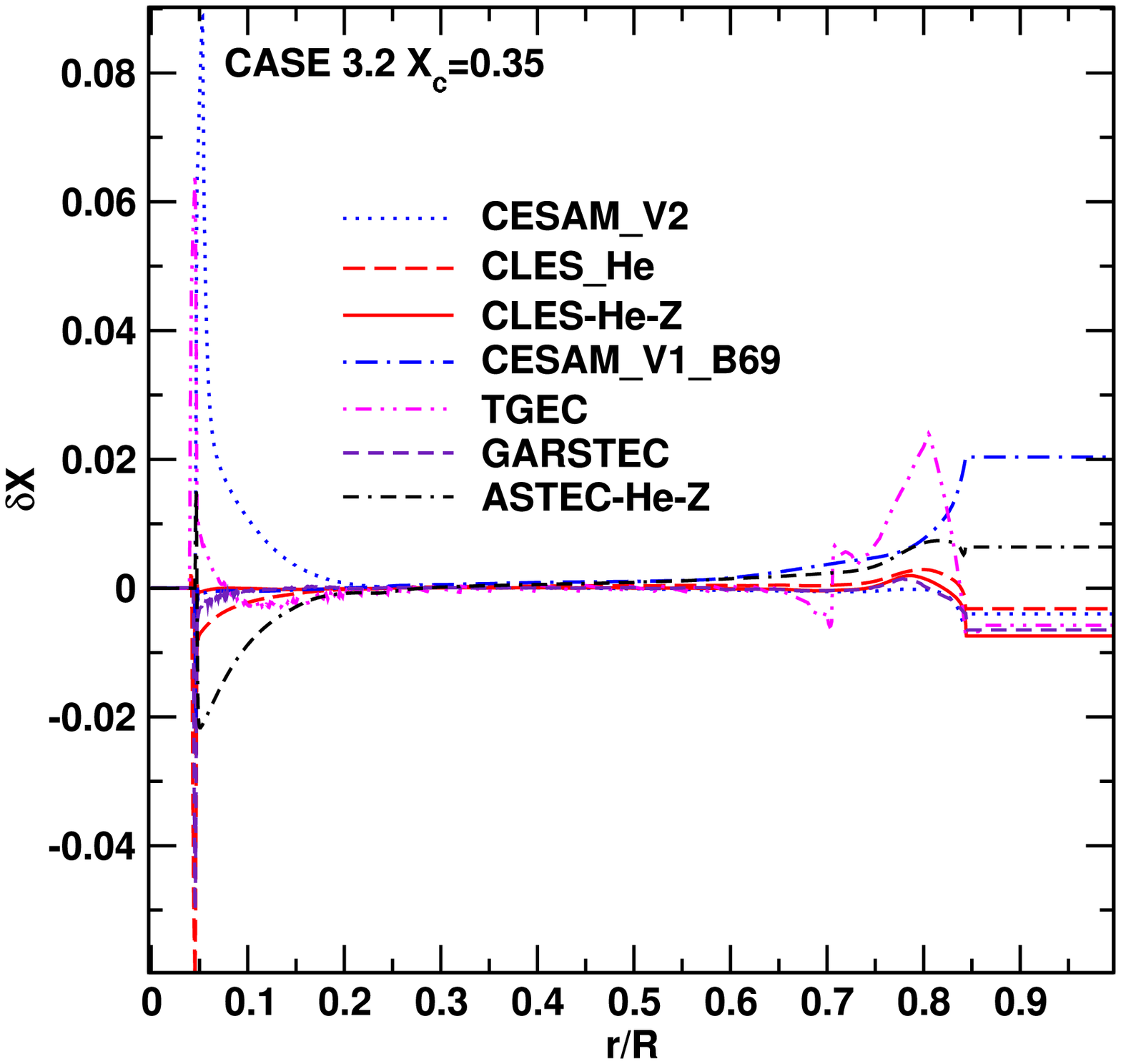}\hspace*{0.4cm}\includegraphics*{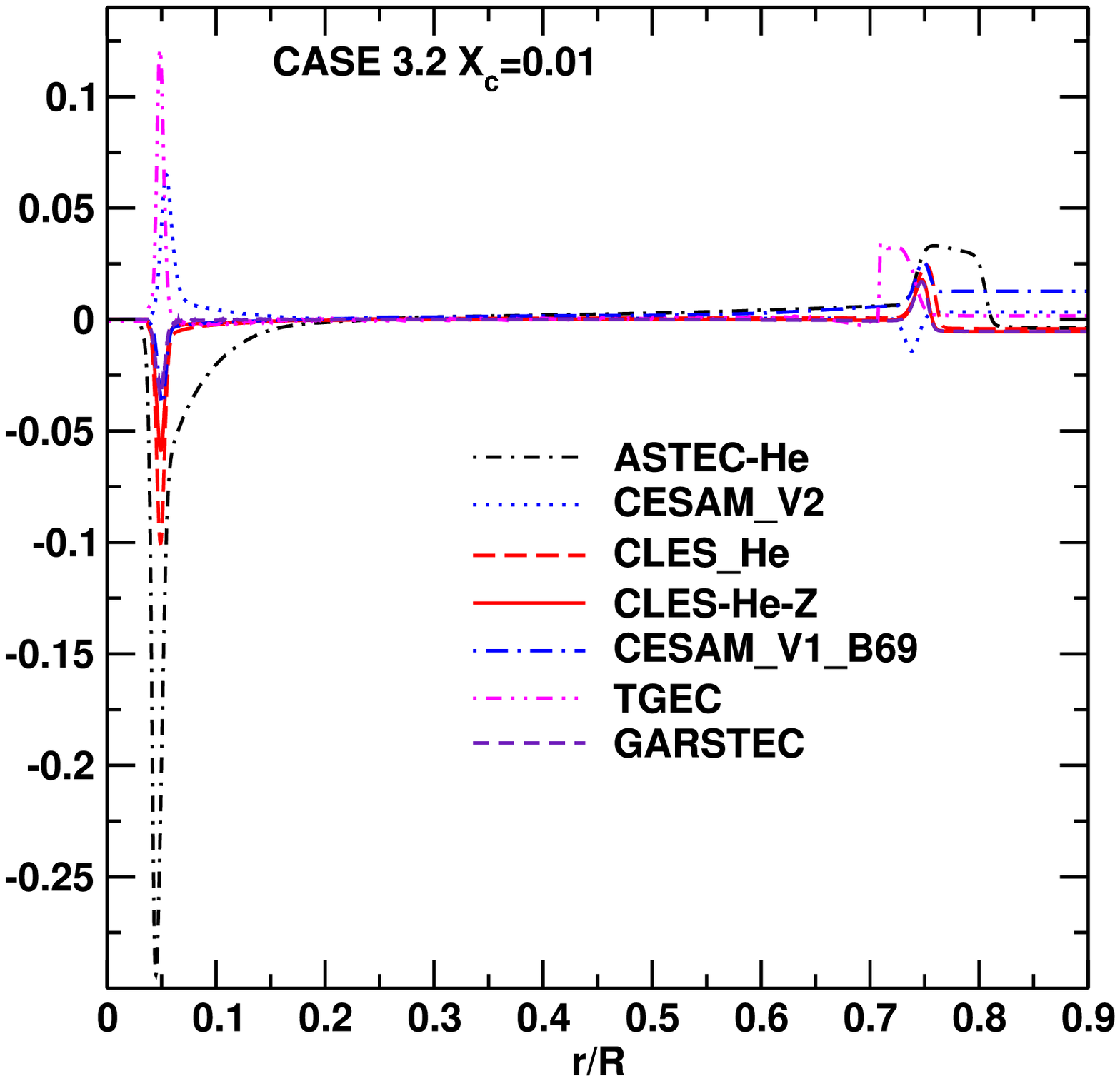}\hspace*{0.7cm}\includegraphics*{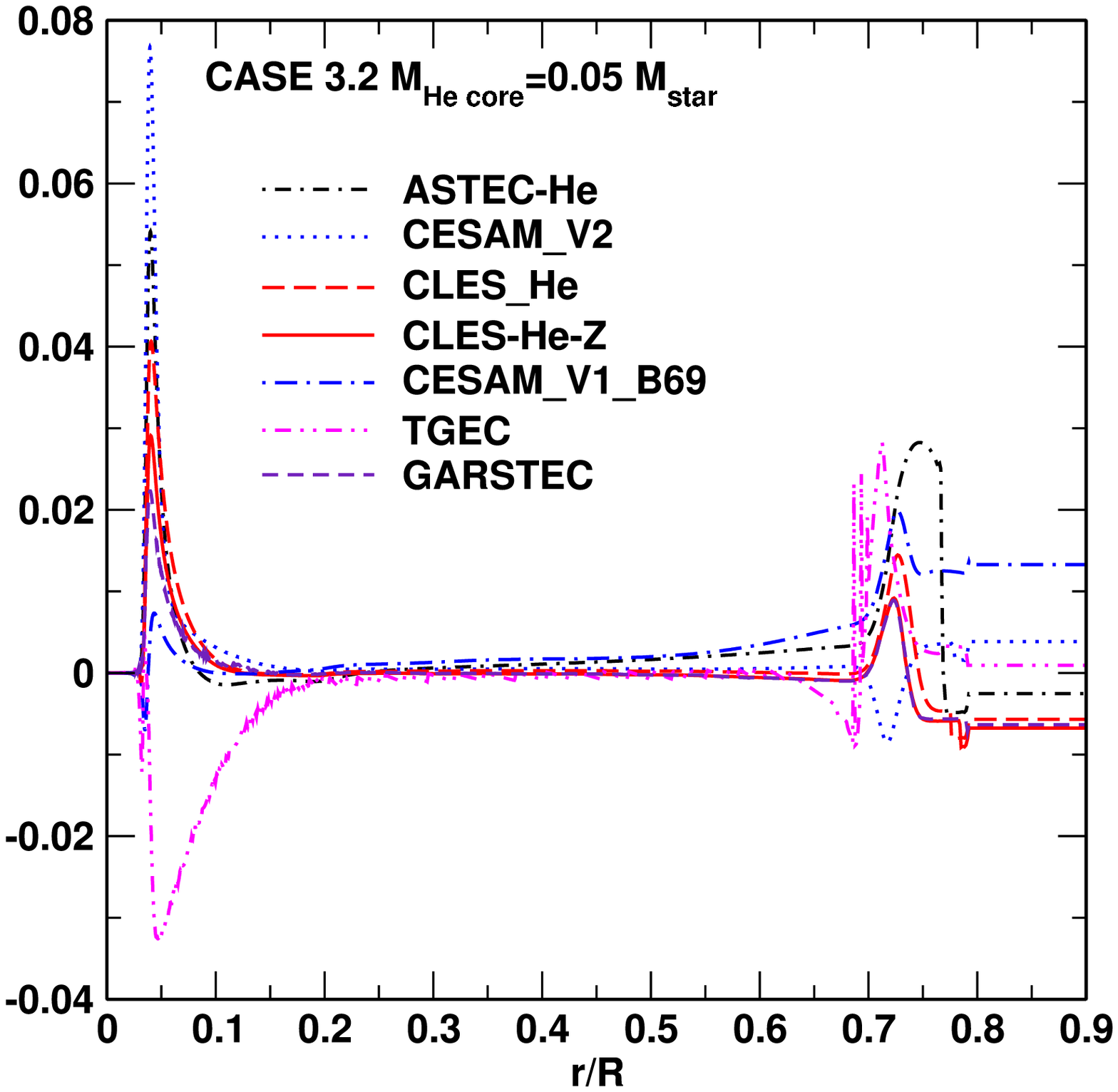}}
\resizebox*{\hsize}{!}{\hspace*{-0.cm}\includegraphics*{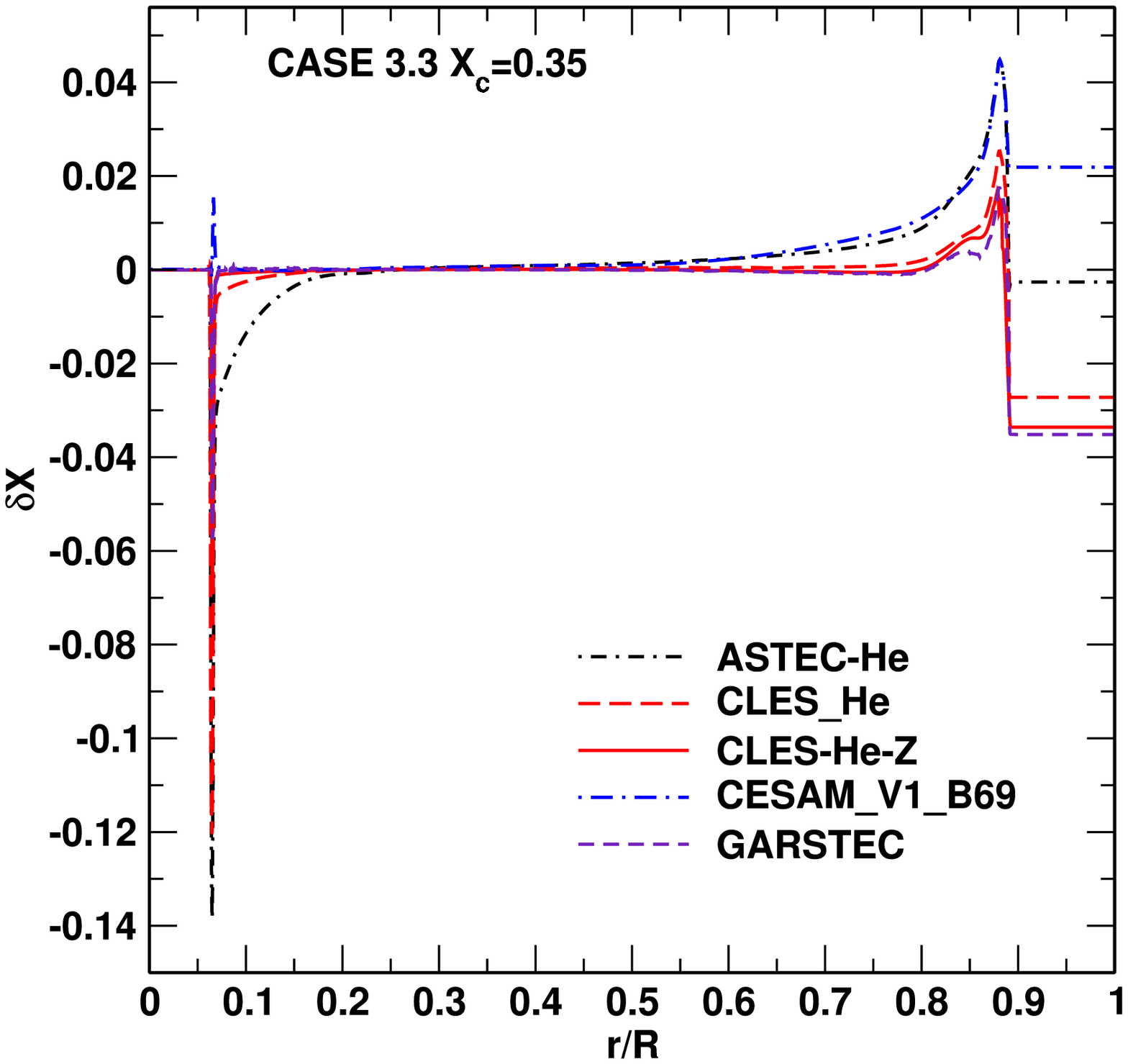}\hspace*{0.4cm}\includegraphics*{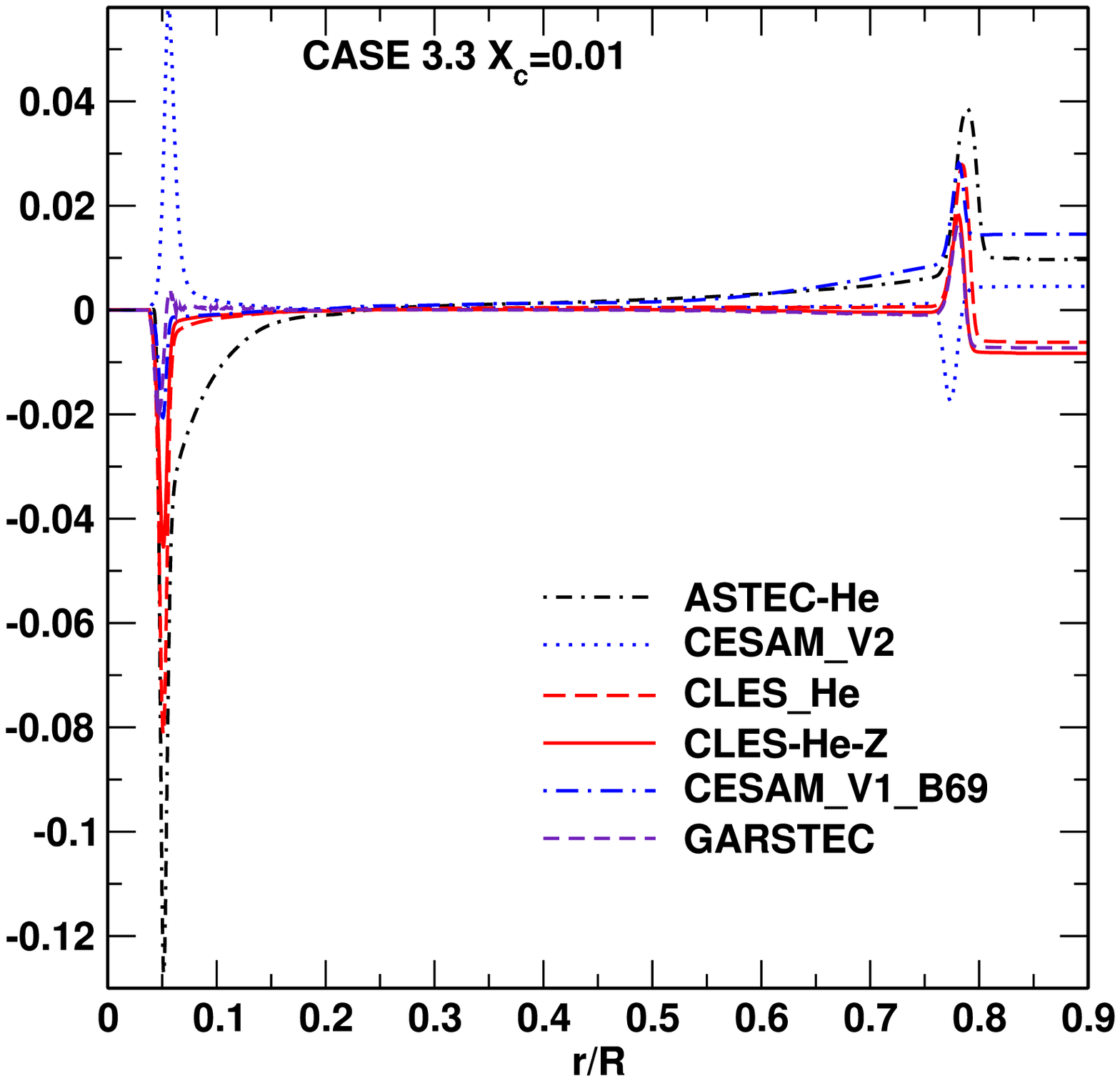}\hspace*{0.7cm}\includegraphics*{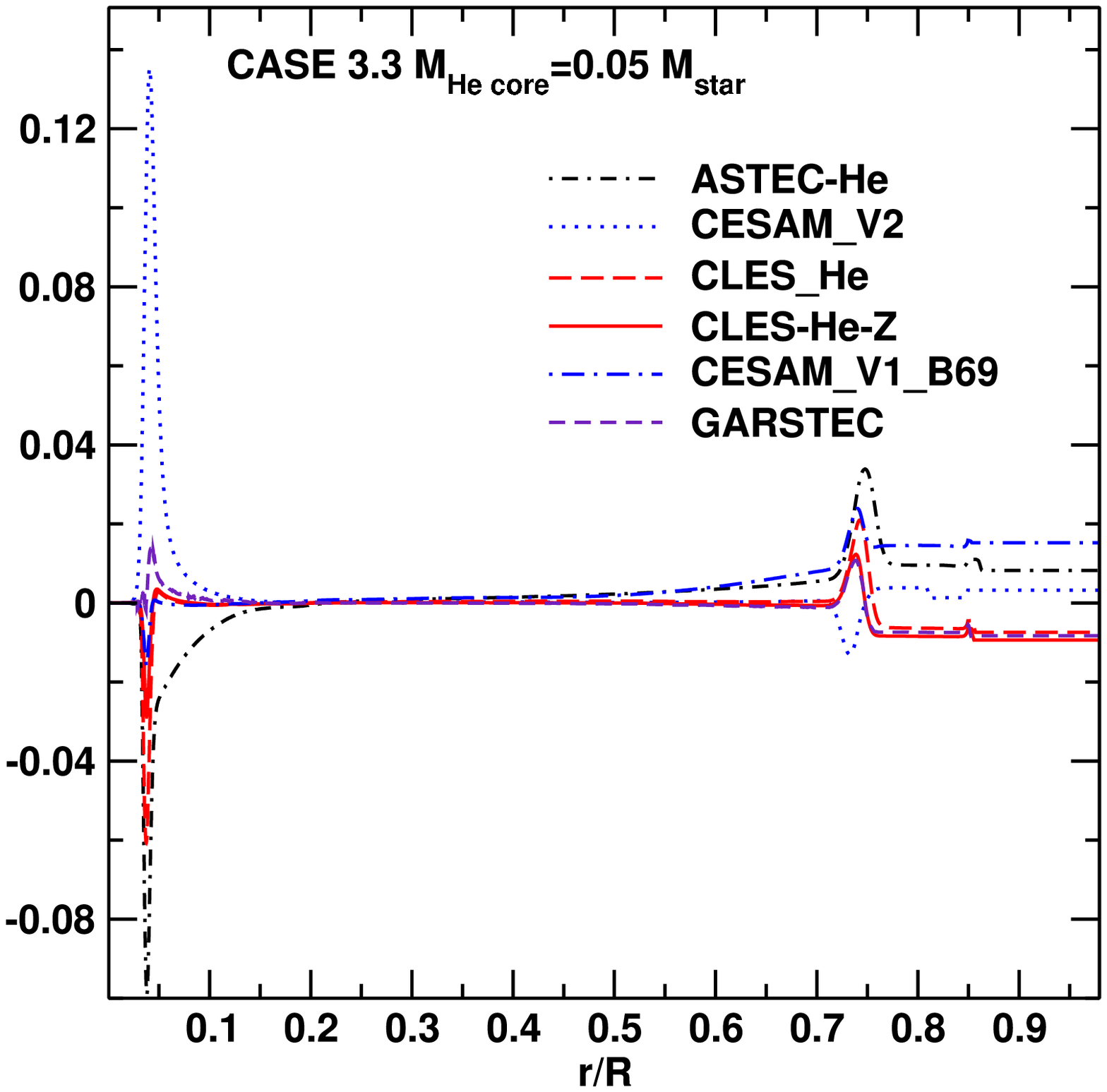}}
\caption{Lagrangian differences of the hydrogen abundance as a function of the normalised stellar radius between each model and the \cesam-V1-MP model corresponding to Case~3.1 (upper panel), Case~3.2 (middle) and Case~3.3 (lower panel) and phases A (left), B (centre) and C (right). The results of different codes or versions of a code have been considered: \astec-He (black, small dot-dash), \astec-He-Z (black, dash-dash-dot), \cesam-V1-B69 (blue, large dot-dash), \cesam-V2 (blue, dots), \cles-He (red, large dash), \cles-He-Z (red, solid line), \garstec\ (indigo, small dash), \tgec\ (magenta, dash-dot-dot).
}
\label{fig:Xc}
\end{center}
\end{figure}

\subsection{Internal structure and surface abundances of the elements}

\begin{figure}[htb]
\begin{center}
\resizebox*{\hsize}{!}{\hspace*{-0.cm}\includegraphics*{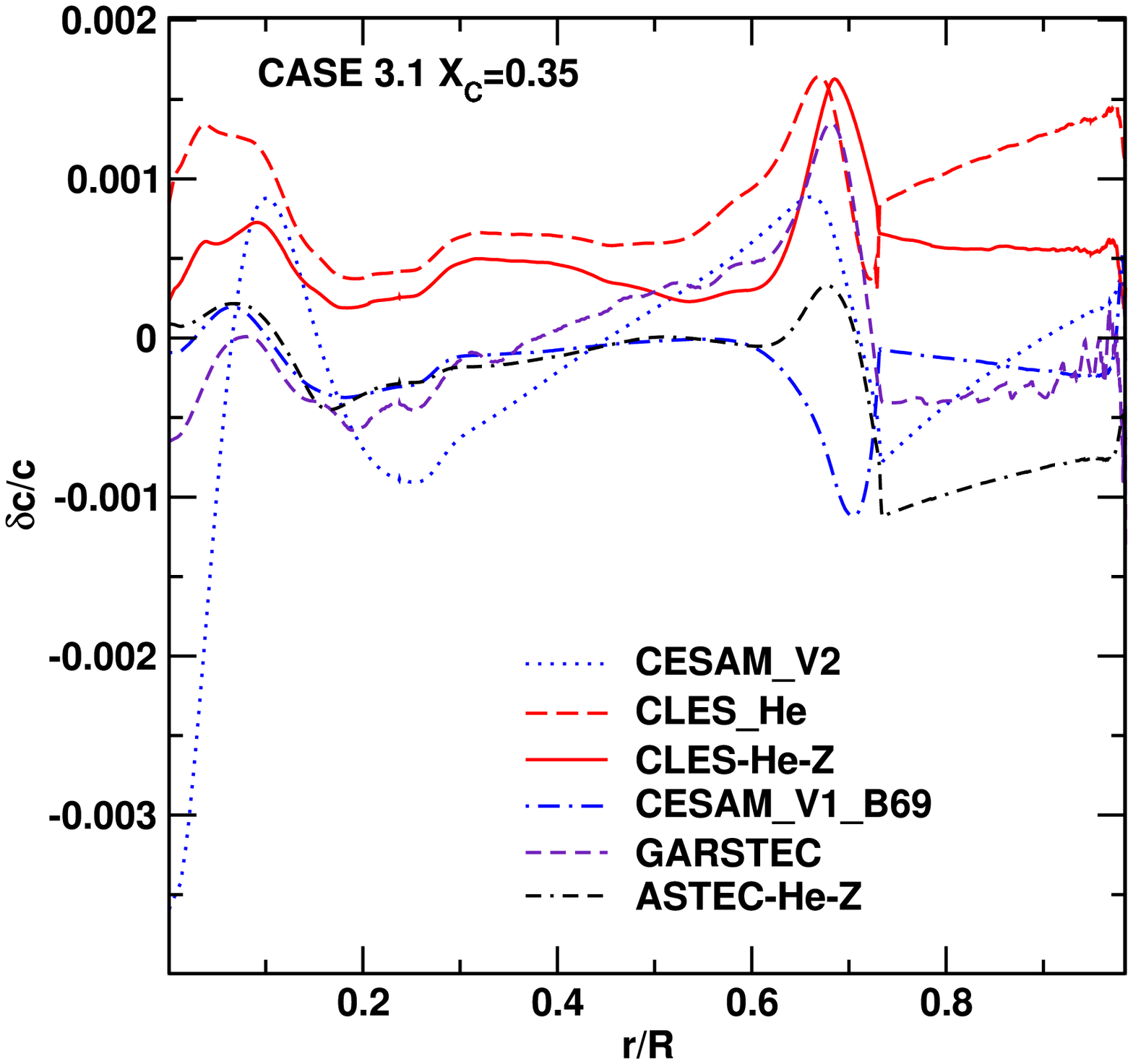}\hspace*{0.4cm}\includegraphics*{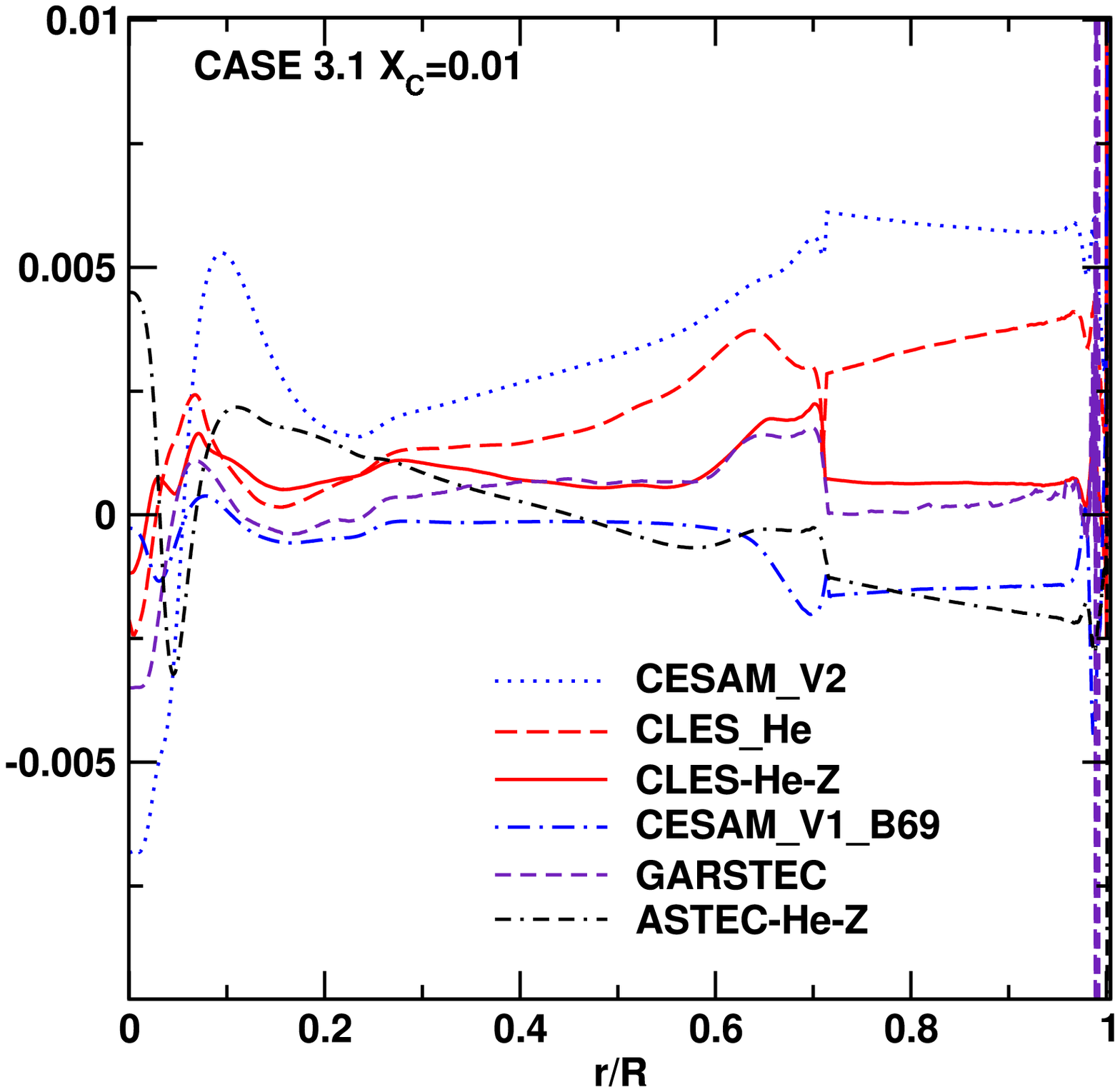}\hspace*{0.7cm}\includegraphics*{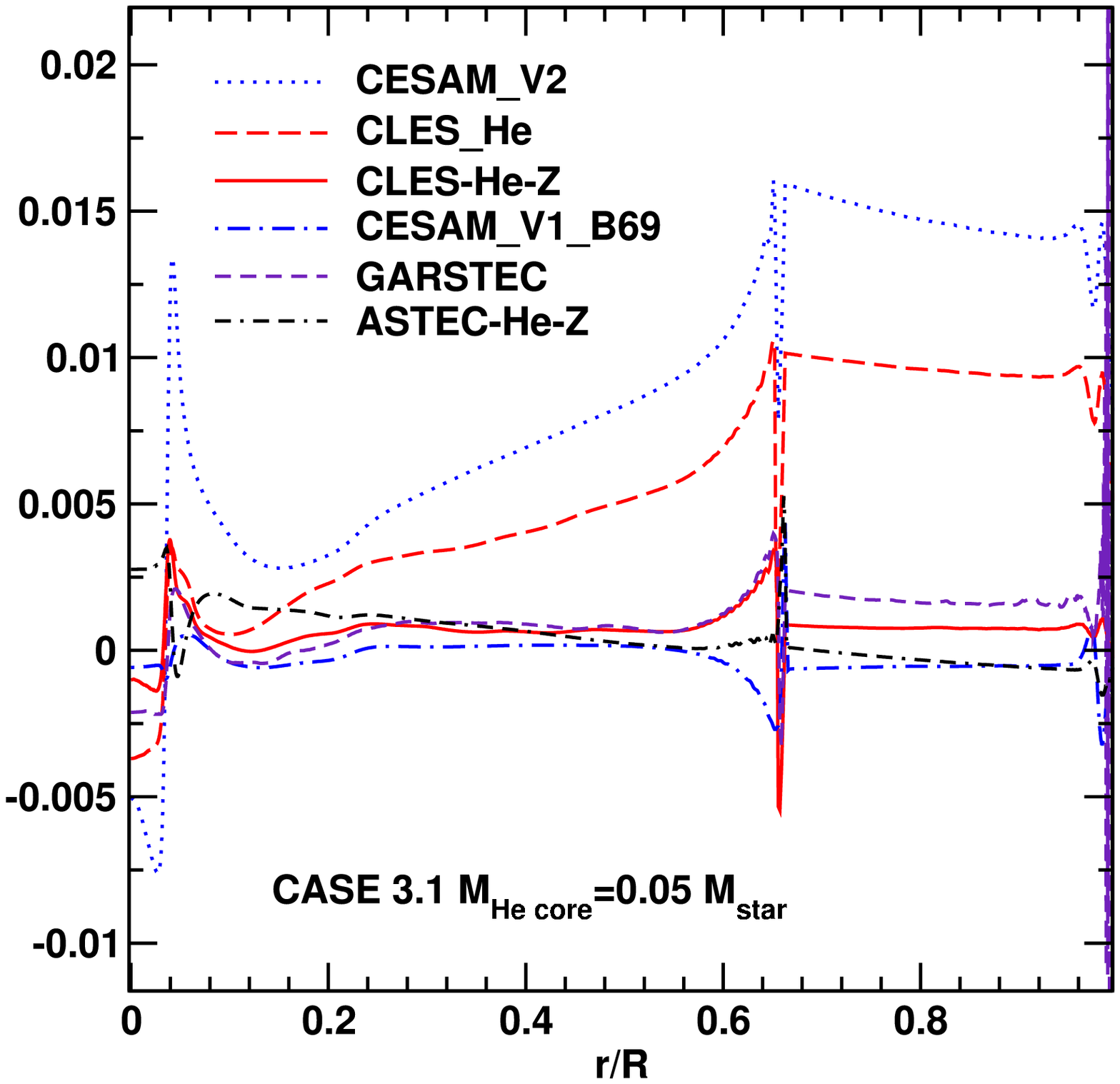}}
\resizebox*{\hsize}{!}{\hspace*{-0.cm}\includegraphics*{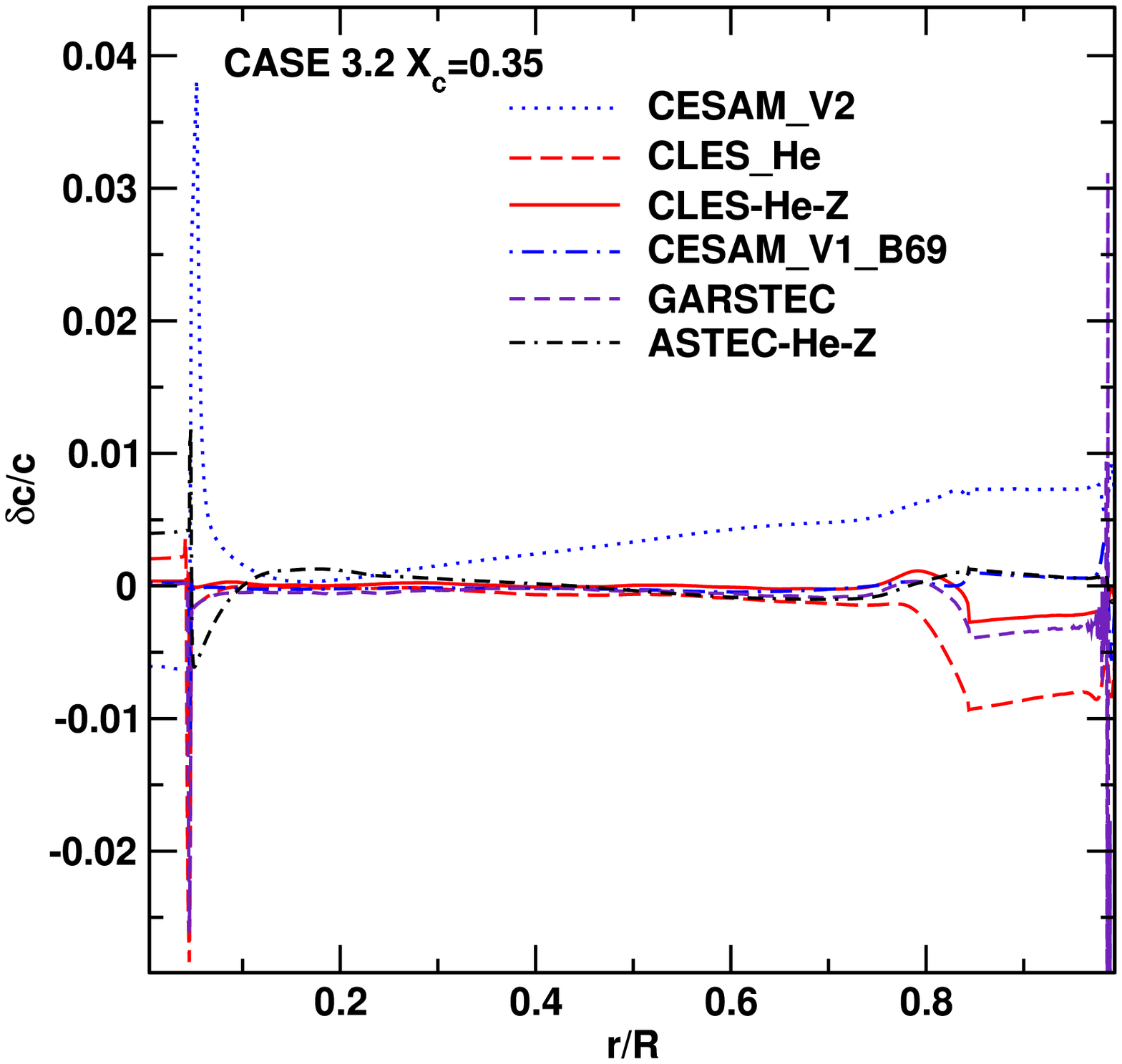}\hspace*{0.4cm}\includegraphics*{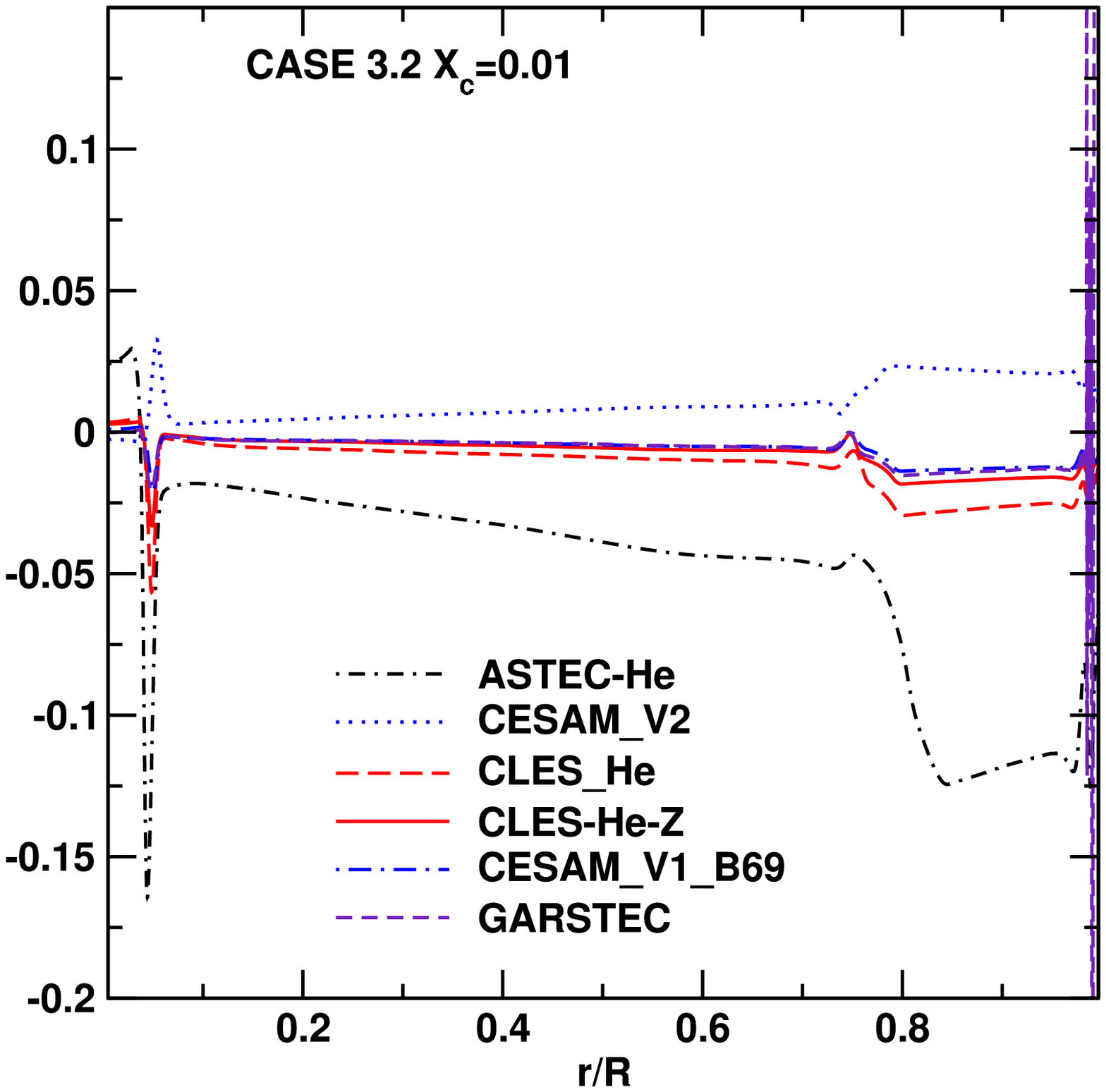}\hspace*{0.7cm}\includegraphics*{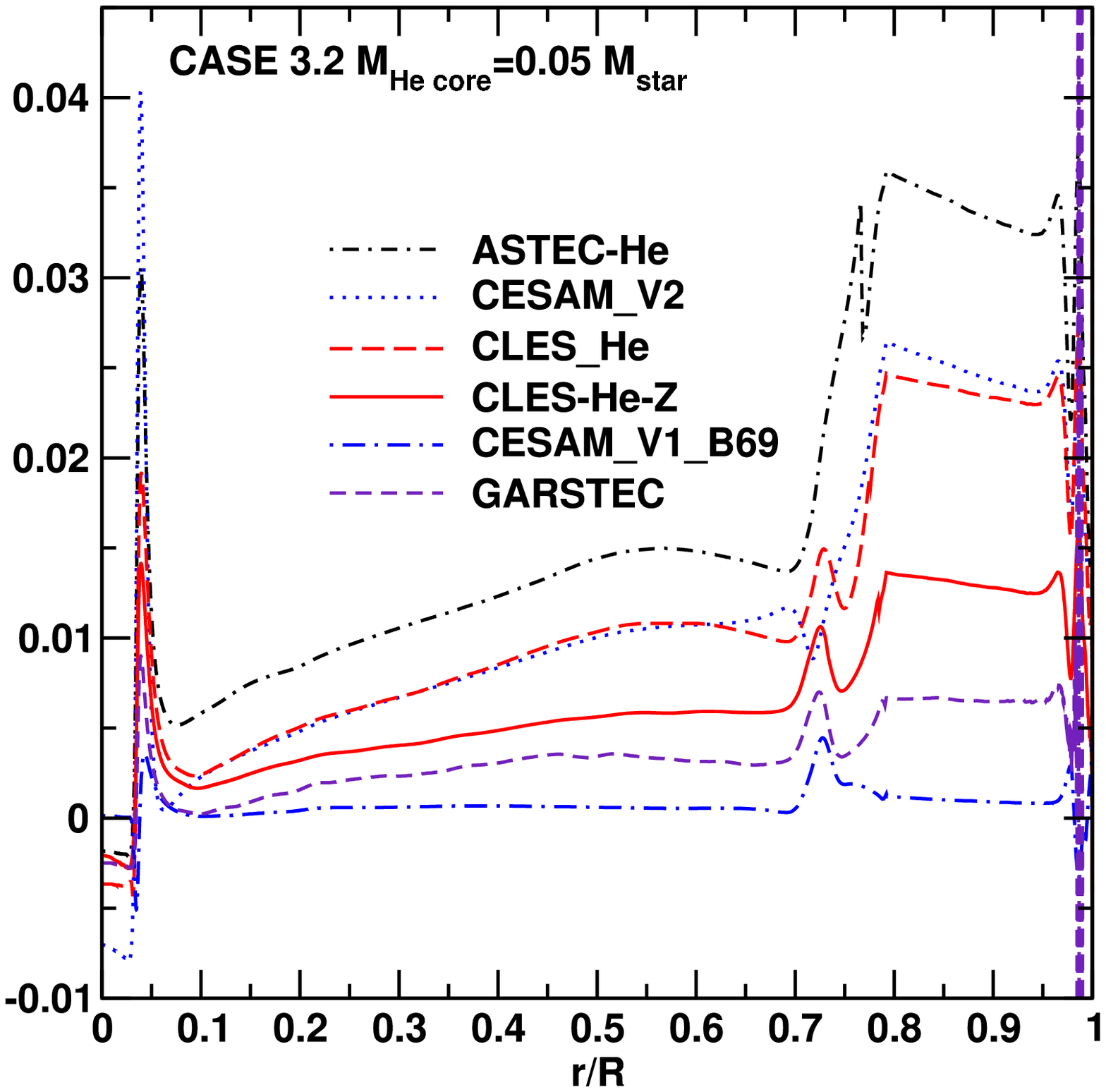}}
\resizebox*{\hsize}{!}{\hspace*{-0.cm}\includegraphics*{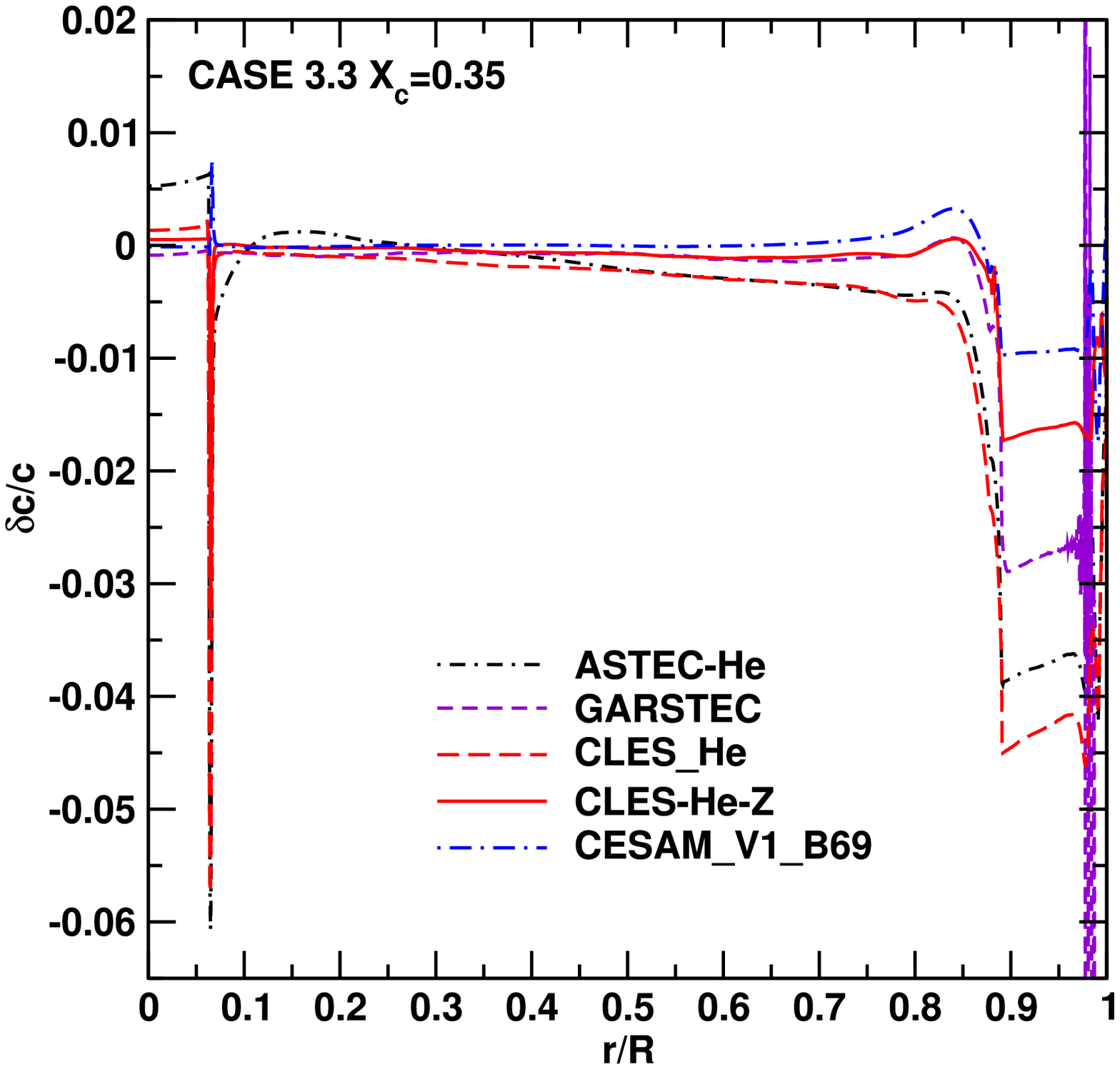}\hspace*{0.4cm}\includegraphics*{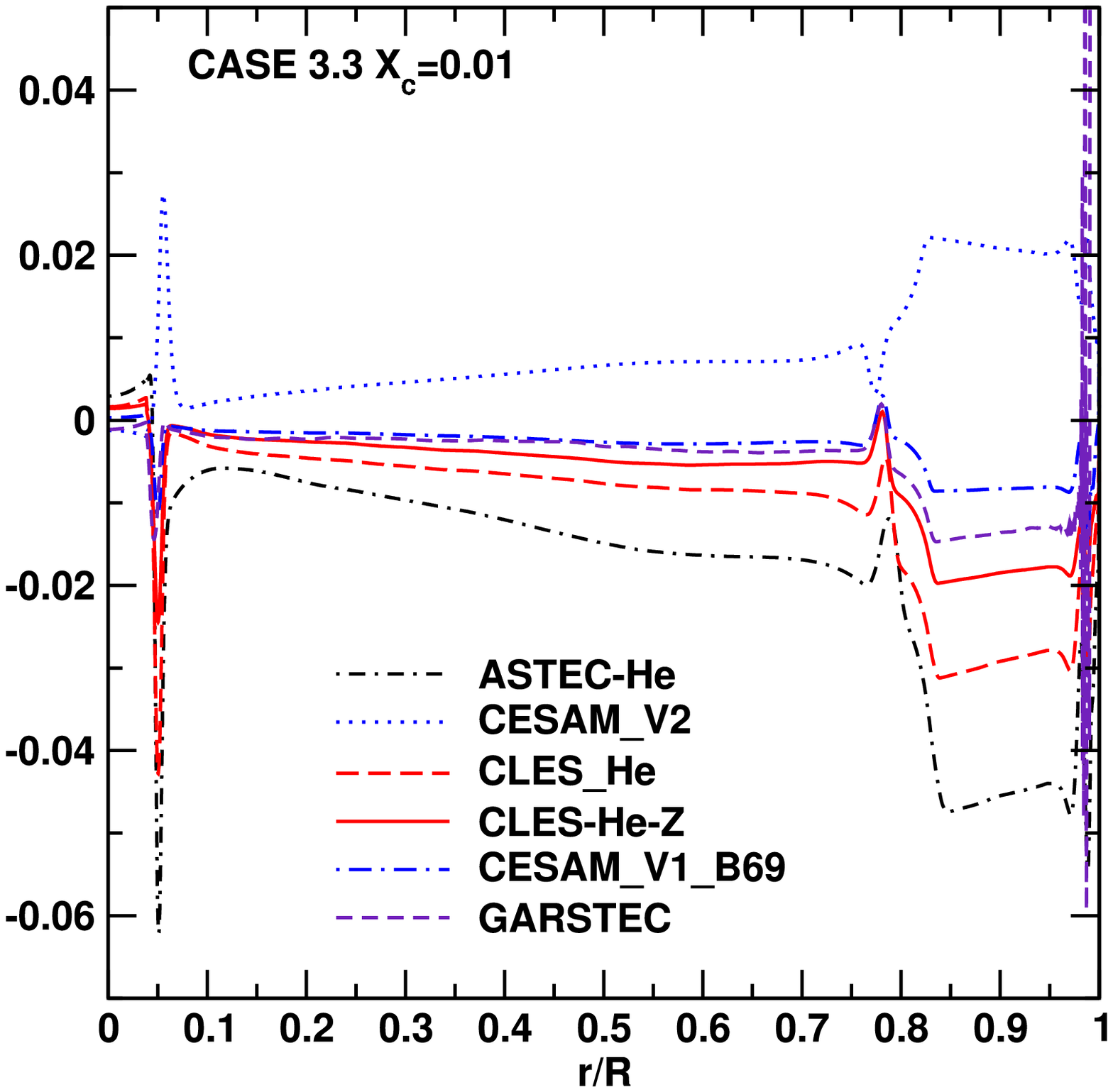}\hspace*{0.7cm}\includegraphics*{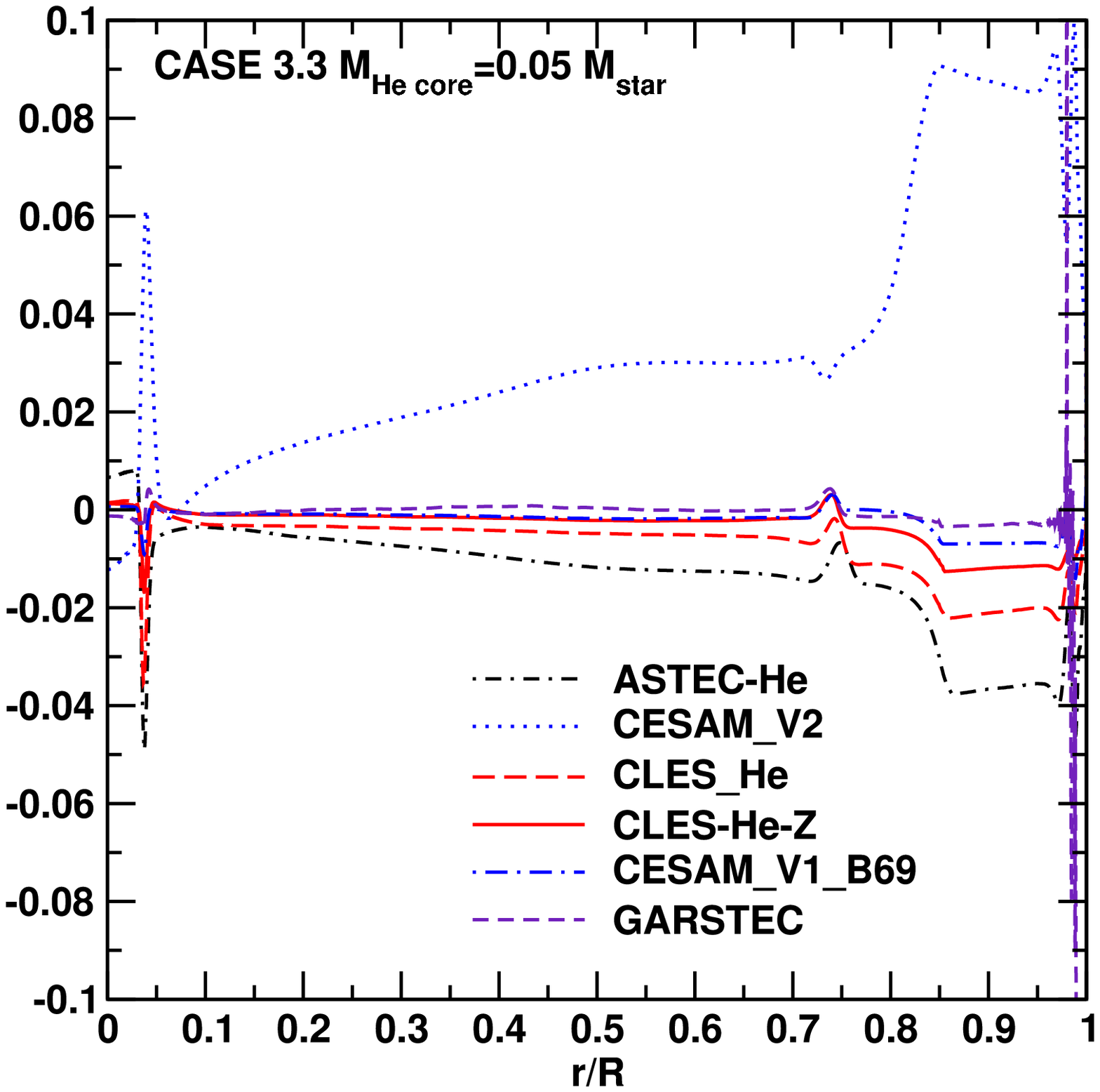}}
\caption{Same as in Fig. \ref{fig:Xc} but for the Lagrangian differences of the sound speed ($\delta \ln c$).}
%\caption{Lagrangian differences of the sound speed ($\delta \ln c$) as a function of the normalised stellar radius between pairs of models corresponding to Case~3.1 (upper panel), Case~3.2 (middle) and Case~3.3 (lower panel) and phases A (left), B (centre) and C (right).
%The results of different codes or versions of a code have been considered: \astec-He (black, %small dot-dash), \astec-He-Z (black, dash-dash-dot), \cesam-V1-B69 (blue, large dot-dash), \cesam-V2 (blue, dots), \cles-He (red, large dash) , \cles-He-Z (red, solid line), \garstec\  (indigo, small dash), \tgec\ (magenta, dash-dot-dot).}
\label{fig:sound}
\end{center}
\end{figure}

\begin{figure}[htb]
\begin{center}
\resizebox*{\hsize}{!}{\hspace*{-0.cm}\includegraphics*{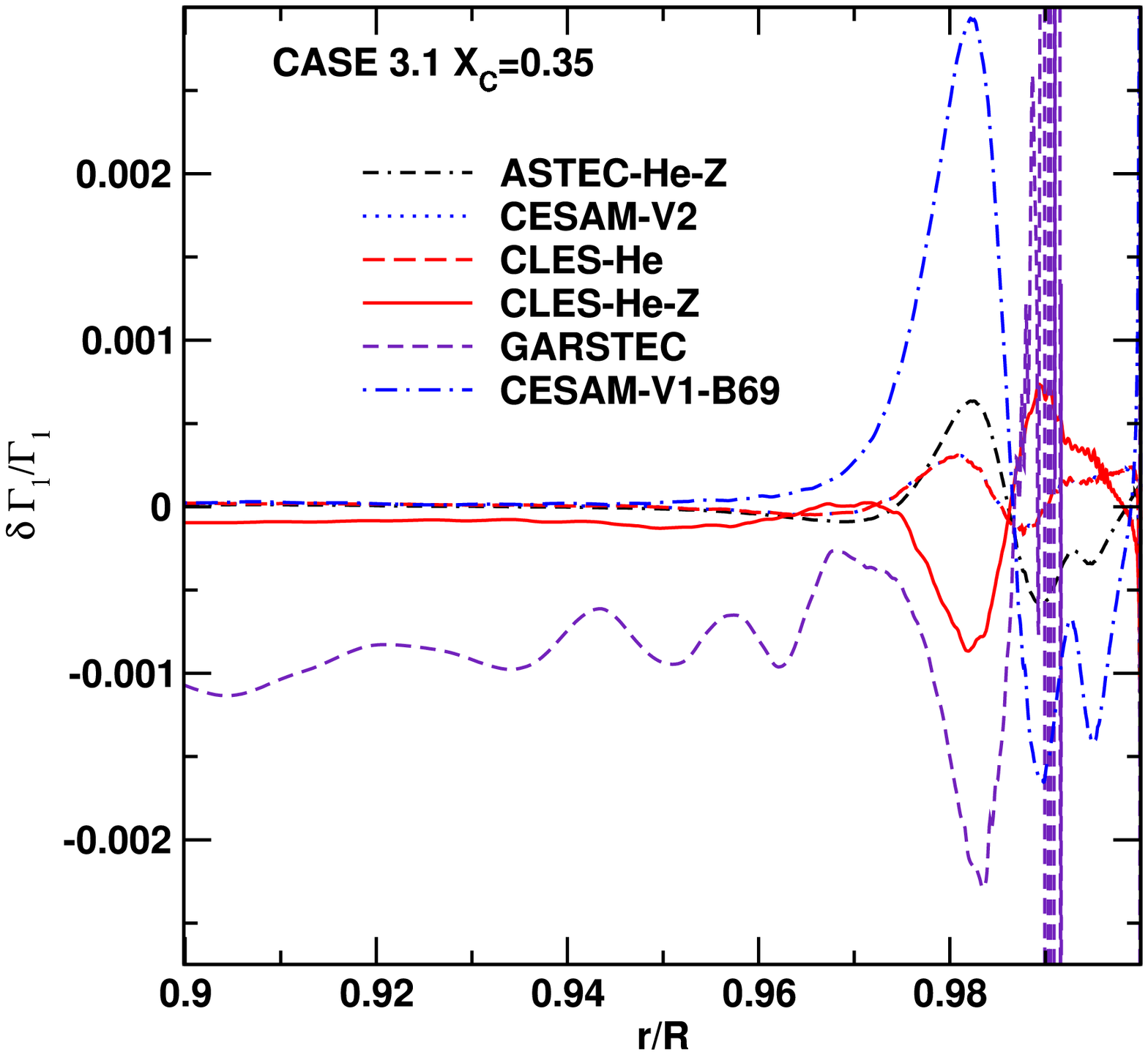}\hspace*{0.7cm}\includegraphics*{./figs/G1_C31C}}
%\resizebox*{\hsize}{!}{\hspace*{-0.cm}\includegraphics*{./figs/G1_C32A}\hspace*{0.7cm}\includegraphics*{./figs/G1_C32C}}
\resizebox*{\hsize}{!}{\hspace*{-0.cm}\includegraphics*{./figs/G1_C33A}\hspace*{0.7cm}\includegraphics*{./figs/G1_C33C}}
\caption{Lagrangian differences of the adiabatic exponent ($\delta \ln \Gamma_1$) as a function of the normalised stellar radius between each model and the \cesam-V1-MP model corresponding to Case~3.1 (upper panel) and Case~3.3 (lower panel) and phases A (left) and C (right). Different codes or versions of a code have been considered: \astec-He (black, small dot-dash), \astec-He-Z (black, dash-dash-dot), \cesam-V1-B69 (blue, large dot-dash), \cesam-V2 (blue, dots), \cles-He (red, large dash), \cles-He-Z (red, solid line), \garstec\  (indigo, small dash).}
\label{fig:gam1}
\end{center}
\end{figure}

Fig.~\ref{fig:Xc} shows the Lagrangian difference $\delta X$ of the hydrogen abundance between each model and the \cesam-V1-MP model calculated at the same mass by means of the \small{ADIPLS} package tools\footnote{http://astro.phys.au.dk/jcd/adipack.n} and plotted as a function of the normalised radius for the 9 models selected (3 cases, 3 phases). Similarly, Figs.~\ref{fig:sound} and \ref{fig:gam1} show, respectively, the Lagrangian difference $\delta \ln c$ of the sound speed and $\delta \ln\Gamma_{1}$  of the adiabatic exponent between each model and the \cesam-V1-MP model. All models provided in the required GONG format have been represented (but see comparisons between \franec\ and \cesam\ in Marconi \cite{marconi07}). Note that the scale on the vertical axis differs from one figure to the other. 

Concerning the hydrogen profile (Fig.~\ref{fig:Xc}), the major differences are located (1) at the frontier of and inside the convective envelope and, (2) in the central regions, i.e. in the region where $r/R_{\mathrm{star}}\lesssim0.25$ for Case 3.1 and at the border of the convective core for Cases 3.2 and 3.3. Differences generally grow as the evolution proceeds from phase A to phase B and C and as mass increases. The differences found are, as expected, larger than those obtained in Task~1 where we compared (simpler) models with no diffusion. The larger differences are generally obtained for the \cesam-V2 and \tgec\ codes where some of the input physics probably do not follow exactly the specifications proposed for the comparisons. We also find rather marked differences of the \astec-He-Z and/or \astec-He models in the central regions mainly for phase B models for the 3 cases.

As discussed by Montalb\`an \etal\ (\cite{mtl07}, hereafter MTL07), the differences in the sound speed in Fig.~\ref{fig:sound} reflect the differences (1) in the radius of the models, (2) in the chemical composition gradient in the envelope for $r>0.6\ R_{\mathrm{star}}$ and, (3) in the position of the boundaries of convection zones. The differences in the adiabatic exponent $\Gamma_{1}$ in the external regions of the model shown in Fig.~\ref{fig:gam1} are related to the difference in the helium abundance in the convective envelope.

\begin{figure}[htb]
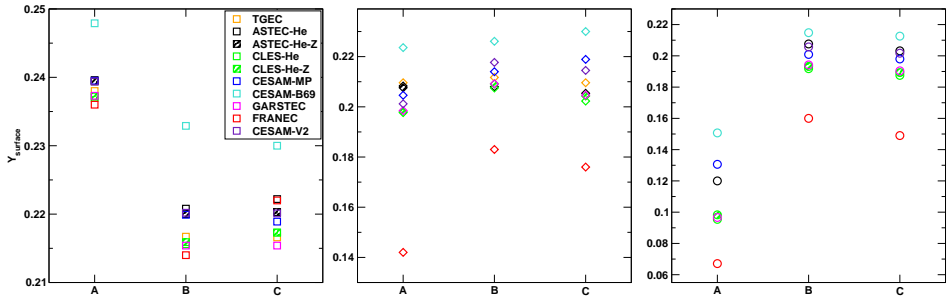

\begin{center}
\resizebox*{\hsize}{!}{\hspace*{-0.cm}\includegraphics*{./figs/Ys1_0}\hspace*{0.4cm}\includegraphics*{./figs/Ys1_2}\hspace*{0.7cm}\includegraphics*{./figs/Ys1_3}}
\caption{Surface abundance of helium in models computed by different evolution codes for Case~3.1 (squares, left), Case 3.2 (diamonds, middle) and Case 3.3 (circles, right) and phases A, B and C. The results of different codes or versions of a code have been considered: \astec-He (black), \astec-He-Z (black-hashed), \cesam-V1-MP (blue), \cesam-V1-B69 (cyan), \cesam-V2 (indigo), \cles-He (green), \cles-He-Z (green-hashed), \franec\ (red), \garstec\ (magenta), \tgec\ (orange).}
\label{fig:Ys}
\end{center}
\end{figure}

The enhancement of the surface abundance of hydrogen and the concomitant depletion of helium and metals differ from one code to another. The surface abundance of He obtained by the different codes is displayed in Fig.~\ref{fig:Ys}, for each case and phase considered. The most important scatter is found for Case~3.1 where differences of the He surface abundance reach 0.07 in phases B and C if \cesam-V1-B69 models are excluded from the comparison and 0.19 if they are included. For Cases 3.2 and 3.3 the differences are lower: they are in the range 0.01-0.03 if \cesam-V1-B69 and \franec\ models are excluded and 0.02-0.08 if they are included. From Fig.~\ref{fig:Ys}, it is clear that in \franec\ models the depletion of the surface helium abundance has a tendency to be larger than in other models while it is always lower in \cesam-V1-B69 models than in other models.

Differences in the diffusion velocities may explain the differences in the surface abundances seen in Fig.~\ref{fig:Xc} and \ref{fig:Ys} (see MTL07 for comparisons between \cles\ and \cesam). Furthermore, because of diffusion metals pile up beneath the convective envelope and the increase of the metal abundance induces an increase of opacity which may trigger convective instability. As discussed by MTL07, this occurs in Cases 3.2 and 3.3 where semiconvection takes place beneath the convective envelope. The evolution of the unstable layers then depends on the numerical treatment of the convective boundaries in the codes. MTL07 show that because of a different treatment of the convection borders, the \cles\ and \cesam\ codes produce external convective zones with different depths which in turn affects the surface abundances.

As discussed by MTL07, semiconvection can also take place at the border of the convective core in Cases 3.2 and 3.3. At the boundary of the convective core, nuclear burning builds a helium abundance gradient. In the diffusion equation, two terms are therefore in competition: the He gravitational settling term which makes helium travel towards the centre and the term due to the composition gradient which pushes helium outwards (Richard \etal\ \cite{richard01}). In Cases 3.2 and 3.3, when the second term becomes dominant, helium goes out of the core and so do the metals and this prevents settling. The increase of the metals at the border of the convective core makes the opacity increase and semiconvection appears.  Again, the different numerical treatments of the convective boundaries in the codes can explain the differences seen in the models in the regions just above the convective core in Cases 3.2 and 3.3 (see Fig.~\ref{fig:Xc}).

\section{Conclusion}

In \esta-Task~3 we have compared models of $1.0, 1.2, 1.3$\ \msol\ on the M--S and subgiant branch which were computed with six different stellar evolution codes and different implementations of microscopic diffusion. 

We found rather large differences in the H--R diagram and internal structure for the \tgec, \franec\ and to a lesser extent \cesam-V2 models. These differences are probably due to small remaining differences between the basic input physics of these codes and those specified for Task~1 and 3. We also find rather marked differences of the hydrogen abundance profile in the central regions for the \astec\ models in phase B models close to the end of the M--S.

Furthermore we showed that the surface depletion of helium due to diffusion is stronger in \franec\ models and much less strong in \cesam-V1-B69 models compared to the other models. The surface depletion of He is sensitive to the  numerical treatment of the convective borders, particularly in presence of semiconvection just beneath the convective envelope and to the diffusion velocities. A further step in the comparison will consist in examining in detail the treatment of the convective boundaries in each code. 

The differences between \cesam-V1-B69 models and others are presently not understood. For instance, \cles\ and \cesam\ use different diffusion velocities because \cles\ is based on the TBL94 formalism while \cesam\ uses Paquette \etal\ (\cite{paquette86}) collisions integrals. Furthermore \cesam\ follows explicitly each element inside $Z$ and determines the ionisation degree of each species while some codes, like \cles\ adopt full ionisation and follow a reduced number of species. However, the consequences of these differences were carefully examined by MTL07, who showed that while the differences between \cesam-V1-MP and \cles\ models are rather well understood it is not the case for the differences between \cesam-V1-B69 and \cles\ models. To progress, it will be necessary to go deeper into the tests of the algorithm that solves Burgers' equations in \cesam.

\acknowledgements
The European Helio and Asteroseismology Network (HELAS) is thanked for financial support.
\endacknowledgements
%%-----------------------------
%%      your bibliography
%%-----------------------------


\begin{thebibliography}{99}
\expandafter\ifx\csname natexlab\endcsname\relax\def\natexlab#1{#1}\fi

%% Using \cite{Bei} in the text

\bibitem[2007]{alecian07}
Alecian, G.: 2007, \EAS

\bibitem[1969]{burgers69}
Burgers, J.~M.: 1969, [B69], {\it Flow Equations for Composite Gases},
New York: Academic Press, 1969

\bibitem[1970]{chapman70}
{Chapman}, S. and {Cowling}, T.~G., 1970, [CC70], {\it The mathematical theory of non-uniform gases. an account of the kinetic theory of viscosity, thermal conduction and diffusion in gases}, Cambridge: University Press, 1970, 3rd ed.

\bibitem[2007]{jcd07}
Christensen-Dalsgaard, J.: 2007, \workshop

\bibitem[2007]{jcddm07}
Christensen-Dalsgaard, J., Di Mauro M.P.: 2007, \EAS

\bibitem[2005]{scilla05}
Degl'Innocenti, S., Marconi M.: 2005, in Joint {\small ENEAS}-\corot/\esta\  Workshop, Aarhus, Denmark at
  {\tt\small http://www.astro.up.pt/corot/welcome/meetings/m4/}

\bibitem[2007]{hui07}
Hui Bon Hoa, A., 2007, \workshop

\bibitem[2006]{lebreton06}
  {Lebreton} Y., 2006, \corot Week 10, \corot/\esta Meeting 6, Nice - France, at 
  {\tt\small http://www.astro.up.pt/corot/welcome/meetings/m6/}

\bibitem[2007]{lebreton07}
{Lebreton} Y., 2007, \workshop
  
\bibitem[2007]{marconi07}
Marconi, M.: 2007, \workshop

\bibitem[2007]{mathis07}
Mathis, S.: 2007, \EAS

\bibitem[1993]{mp93}
Michaud, G. and Proffitt, C.~R.: 1993, [MP93], in W.~W. Weiss and A. Baglin (eds.), {\it ASP Conf. Ser. 40: IAU Colloq. 137: Inside the Stars}, pp 246--259

\bibitem[2007]{mtl07}
Montalb\`an, J., Th\'eado, S., Lebreton, Y.: 2007, [MLT07], \EAS

\bibitem[2006]{monteiro06}
Monteiro, M.~J.~P.~F.~G., Lebreton, Y., Montalb\`an, J.,
  Christensen-Dalsgaard, J., Castro, M., Degl'Innocenti, S., Moya., A.,
  Roxburgh, I.~W., and Scuflaire, R. et \etal: 2006,
in F. Favata, A. Baglin, and J. Lochard (eds.), {\it ESA Publications
  Division, ESA SP; ESA Spec.Publ. 1306}, pp 363--372

\bibitem[1997]{morel97}
{Morel} P., 1997, A\&As, 124, 597
  
\bibitem[2007]{moya07}
Moya, A.: 2007, \EAS

\bibitem[1986]{paquette86}
Paquette, C., Pelletier, C., Fontaine, G., and Michaud, G.: 1986,
{ApJS} {61}, 177

\bibitem[2001]{richard01}
Richard, O., Michaud, G., Richer, J.: 2001, ApJ 558, 377

\bibitem[2003]{schlattl03}
Schlattl, H. and Salaris, M., 2003, A\&A 402, 29

\bibitem[2007]{theado07}
Th\'eado, S., 2007, \workshop

\bibitem[1994]{tbl94}
Thoul, A.~A., Bahcall, J.~N., and Loeb, A.: 1994, [TBL94], ApJ, 421, 828 

\bibitem[2007]{tm07}
Thoul, A.~A., Montalb\`an, J.: 2007, \EAS

\bibitem[1998]{turcotte98}
Turcotte, S., Richer, J., and Michaud, G.: 1998, ApJ, 504, 559

\bibitem[2005]{weiss05}
  {Weiss} A., 2005, in \corot/\esta Meeting 3, Nice, France, at
  {\tt\small http://www.astro.up.pt/corot/welcome/meetings/m3/}

\bibitem[2007]{zahn07}
Zahn, J.-P.: 2007, \EAS

\end{thebibliography}
\end{document}